    \documentclass[preprint,12pt]{elsarticle}
    
\usepackage{pgfplots}
\pgfplotsset{compat=1.18}
\definecolor{steelcyan}{RGB}{70,130,180}
\definecolor{mygray}{RGB}{89,89,89}
\definecolor{mypink}{RGB}{255,191,204}
\definecolor{mycyan}{RGB}{51,179,230}
    \usepackage{graphicx}
    \usepackage{epstopdf, epsfig}

    \usepackage{amstext}
    \usepackage{amsmath}
    \usepackage{amsfonts,amssymb}
    \usepackage{framed}
    \usepackage{algorithm}
    \usepackage{algorithmicx}
    \usepackage{algcompatible}
    \usepackage{subcaption}
    \usepackage{lipsum}
    \usepackage{bm}
    \usepackage{xcolor}
    \usepackage{pdfpages}
    \usepackage{lineno}
    \usepackage{mathpazo}
    \usepackage{calrsfs}
    \DeclareMathAlphabet{\pazocal}{OMS}{zplm}{m}{n}

    \usepackage{amssymb}
    \usepackage{amsmath}
    \usepackage{natbib}
    \usepackage{epstopdf}
    \usepackage{algorithm}
    \usepackage{algpseudocode}
    \usepackage{booktabs} 
    \usepackage{multirow} 
    \journal{}
    
\begin{document}
    
    \begin{frontmatter}
    
    
    
    \title{Impact of Wave Interference on the Consistency Relations of Internal Gravity Waves near the Ocean Bottom}
    
    
    \author[label1]{Guangyao Wang\corref{cor1}} 
    \author[label1]{Yue Wu}
    \author[label1]{Yulin Pan} 
    \author[label2]{Kayhan Momeni} 
    \author[label3]{Joseph Skitka} 
    \author[label4]{Dimitris Menemenlis}
    \author[label5]{Brian K. Arbic}
    \author[label2]{William R. Peltier}
    \cortext[cor1]{Corresponding author.
    E-mail address: guangyaowang1990@gmail.com}
    \affiliation[label1]{organization={Department of Naval Architecture and Marine Engineering, University of Michigan},
                city={Ann Arbor, MI}, country={USA}}
    \affiliation[label2]{organization={Department of Physics, University of Toronto},
                city={Toronto, ON},
                country={Canada}}
    \affiliation[label3]{organization={Woods Hole Oceanographic Institution},
                city={Woods Hole, MA},
                country={USA}}

    \affiliation[label4]{organization={Moss Landing Marine Laboratories, San Jose State University},
                city={Moss Landing, CA},
                country={USA}}
    \affiliation[label5]{organization={Department of Earth and Environmental Sciences, University of Michigan},
                city={Ann Arbor, MI}, country={USA}}

\begin{abstract}
Consistency relations of internal gravity waves (IGWs) describe ratios of cross-spectral quantities as functions of frequency. It has been a common practice to evaluate the measured or simulated signals (e.g., time series of velocity, density, etc.) against the consistency relations, as a way to determine whether an oceanic field of interest is comprised of IGWs. One such study is carried out in Nelson et al.~({\it{JGR Oceans}}, 125(5), 2020, e2019JC015974), which certifies that the ocean interior field in a numerical simulation of a region southwest of Hawaii is dominated by IGWs, through evaluating the consistency relations derived from time series at a depth of $620~\text{m}$. However, we find that when the same procedure is applied at greater depths (e.g., $2362~\text{m}$,~$3062~\text{m}$, and~$4987~\text{m}$), a clear deviation of the simulated signal from the classical consistency relations is observed. In this paper, we identify the reason for the unexpected deviation and show that it is a general phenomenon due to interference of low vertical modes under the reflection by the ocean bottom. We further derive a new set of formulae to characterize the consistency relations of these low modes and validate these formulae using model output.
\end{abstract}
\end{frontmatter}
    
    
\section{Introduction}

\label{sec:intro}
As ubiquitous features of the ocean, Internal Gravity Waves (IGWs) are generated by perturbations of stratified fluids. As these waves propagate, they undergo nonlinear interactions and eventually break down into small-scale turbulence~\citep{schooley1972experimental,lamb2014internal,polzin2011toward,susanto2005ocean}. It is hypothesized and believed by many that IGWs drive the downscale energy cascade in the interior of the ocean, thus serving as the main contributor to ocean mixing~\citep{alford2016near,whalen2020internal}.  

The hypothesis of a downscale energy cascade implies that the ocean interior is filled with a continuous spectrum of IGWs, known as the IGW continuum~\citep{garrett1972space}. Among many efforts to verify the IGW continuum, one commonly used approach is to assess whether the observed or modeled data in the ocean satisfy the dispersion relation or consistency relations of IGWs, e.g.,~\cite{muller1976consistency,lien1992consistency,polzin2011toward,pan2020numerical,nelson2020improved}. In this study, we focus on the consistency relations, which describe the dependence of oscillation amplitudes of velocity, pressure, buoyancy, etc. As an example, for a monochromatic plane wave of frequency $\omega$, the ratio of its vertical kinetic energy $E_{vk}$ to horizontal kinetic energy $E_{hk}$ and the ratio of $E_{hk}$ to potential energy $E_p$ can be formulated as~\citep{lien1992consistency,polzin2011toward},
\begin{eqnarray}
    \frac{E_{vk}}{E_{hk}}&=& \frac{(\omega^2-f^2)\omega^2}{(\omega^2+f^2)N^2},
\label{eq:r11}
\end{eqnarray}
\begin{eqnarray}
    \frac{E_{hk}}{E_p}&=& \frac{\omega^2+f^2}{\omega^2-f^2},
\label{eq:r22}
\end{eqnarray}
where $f$ is the Coriolis frequency, and $N$ is the Brunt–V\"ais\"al\"a or buoyancy frequency. 

\begin{figure}
  \centerline{\includegraphics[trim=0cm 0cm 0cm 0.09cm, clip,scale=4.0]{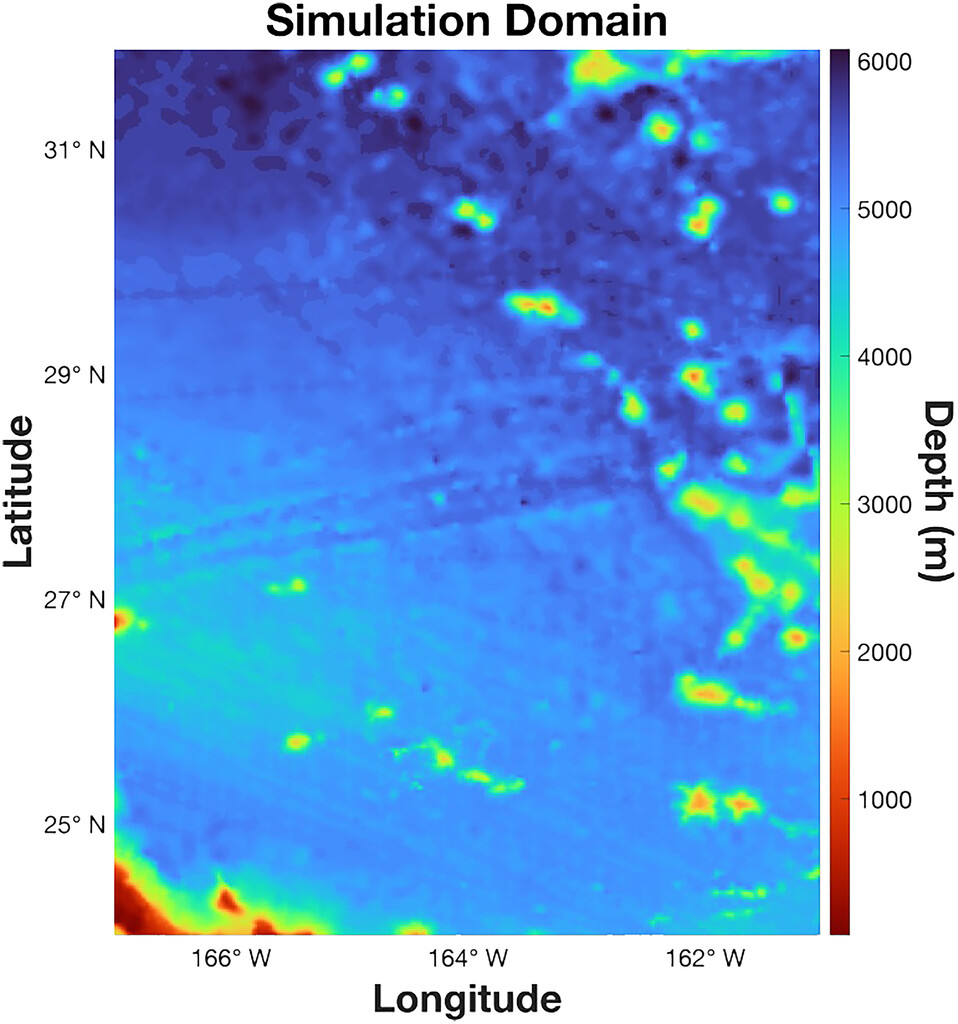}}
  \caption{Bathymetry in the simulation domain (24$^\circ$N to 32$^\circ$N, 193$^\circ$E to 199$^\circ$E). Adapted from \protect\cite{skitka2024probing}.}
\label{fig:domain}
\end{figure}

In~\cite{nelson2020improved}, the authors performed an analysis on the consistency relation using data from a Massachusetts Institute of Technology general circulation model~\citep[MITgcm,][]{marshall1997finite} simulation of a region northwest of Hawaii (Fig.\,\ref{fig:domain}), with the initial conditions taken from, and the lateral boundaries forced by the output variables (including temperature, salinity, and velocities) from a global ocean model simulation commonly referred to as LLC4320~\citep{rocha2016seasonality} and atmospheric fields from European Centre for Medium-Range Weather Forecasts (ECMWF). In particular, by analyzing the time series of the zonal velocity $u$, meridional velocity $v$, and vertical velocity $w$ at a particular spatial point, they computed 
\begin{equation}
    E_{vk}=\frac{\tilde{w} \tilde{w}'}{2},~E_{hk}=\frac{\tilde{u} \tilde{u}'+\tilde{v}\tilde{v}'}{2},
\label{eq:vkhk}
\end{equation}
where~$\tilde{}$~denotes Fourier coefficients and $'$ represents complex conjugate. Upon evaluating against Eq.\,\eqref{eq:r11} using data at $(25^{\circ} \text{N}, 195^{\circ}\text{E})$ and a depth level of $620~\text{m}$, they concluded that Eq.\,\eqref{eq:r11} is well satisfied, especially when the grid resolution becomes higher~\cite[see Fig.\,7 in][]{nelson2020improved}, so that the ocean interior is indeed dominated by the IGW continuum. The same procedure has also been applied to ocean observations in many papers, including an application in \cite{polzin2011toward} to a site (named Site D) north of the Gulf Stream. In addition to $E_{vk}$ and $E_{hk}$, the authors also computed $E_p$ as
\begin{equation}
    E_p=\frac{1}{2}N^{-2}\tilde{b} \tilde{b}',
\label{eq:pe}
\end{equation}
where $b=-g\Delta\rho/\rho_0$ is the buoyancy, with $g$ the gravity acceleration, $\rho_0$ the mean density, and $\Delta\rho$ the deviation of fluid density from $\rho_0$. With all time series obtained at depth levels of either $305~\text{m}$ or $391~\text{m}$, it was shown that both Eqs.\,\eqref{eq:r11} and \eqref{eq:r22} are well satisfied at higher frequencies~\cite[see Fig.\,3 in][]{polzin2011toward}, again supporting the IGW continuum argument.

\begin{figure}
  \centerline{\includegraphics[trim=0cm 0cm 0cm 0.0cm, clip,scale=0.75]{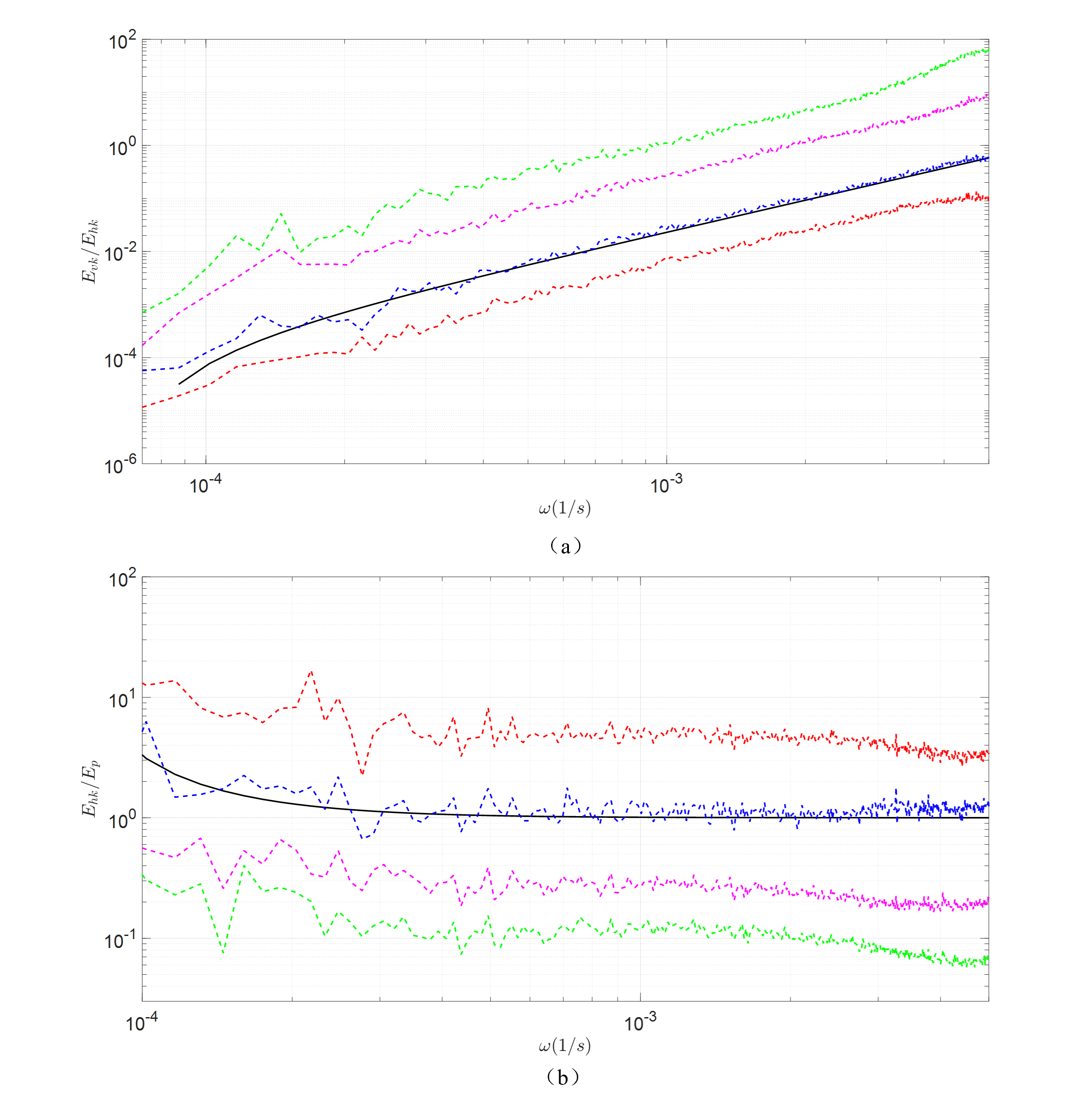}}
  \caption{Results of (a) ${E_{vk}}/{E_{hk}}$ and (b) ${E_{hk}}/{E_{p}}$ at different depths, including $612~\text{m}$ (blue dashed line), $2362~\text{m}$ (red dashed line), $3062~\text{m}$ (magenta dashed line) and $4987~\text{m}$ (green dashed line), in comparison with Eqs.\,\eqref{eq:r11} and \eqref{eq:r22} (black solid line), respectively.}
\label{fig:r1and2}
\end{figure}

While both \cite{nelson2020improved} and \cite{polzin2011toward} suggest that the classical consistency relations, such as Eqs.\,\eqref{eq:r11} and~\eqref{eq:r22}, to be an effective tool in evaluating the IGW continuum, we find some exceptions in our recent research. When the same procedure is repeated at different depth levels in the MITgcm model used in \cite{nelson2020improved}, we find that the favorable agreement found in \cite{nelson2020improved} is only achieved at shallow depth levels. Specifically, Fig.\,\ref{fig:r1and2}(a) shows $E_{vk}/E_{hk}$ evaluated from time series at depths of $612~\text{m}$, $2362~\text{m}$, $3062~\text{m}$, and $4987~\text{m}$. Only at the depth of $612~\text{m}$, similar to that in \cite{nelson2020improved}, is an agreement with the classical consistency relation Eq.\,\eqref{eq:r11} observed. At greater depths, clear deviations (with order-of-magnitude difference) are identified, and these deviations behave in an oscillatory manner as a function of depth. We further note that this phenomenon is general for different classical consistency relations, e.g., see a similar plot for Eq.\,\eqref{eq:r22} in Fig.\,\ref{fig:r1and2}(b). This deviation raises a fundamental question on whether IGWs lose their dominant role at deeper regions or whether the classical consistency relations are  problematic at greater depths.

We show in this paper that the deviations from the classical consistency relations arise from the latter, and that the consistency relations need to be modified when analyzing the time series at a single depth level or spatial point near the ocean bottom. In essence, Eqs.\,\eqref{eq:r11} and~\eqref{eq:r22} are based on a monochromatic plane wave. In applications of the consistency relations to an IGW field, wave interference needs to be considered. The strongest interference occurs for low modes near the ocean bottom, where incident and reflected waves form a standing wave (with node and antinode) vertically. This standing pattern strongly influences the ratios ${E_{vk}}/{E_{hk}}$ and $E_{hk}/E_{p}$ when all $E$'s are evaluated from the time series at a fixed depth level or spatial point. In other words, Eqs.\,\eqref{eq:r11} and~\eqref{eq:r22} should be expected to hold primarily for high modes instead of low modes — a conclusion that may appear counterintuitive, since one might anticipate poorer accuracy at higher modes due to the reduced grid resolution. We show from the model output that this is indeed the case. In particular, when low modes are filtered out from the field, Eqs.\,\eqref{eq:r11} and~\eqref{eq:r22} are well satisfied at any depth. In addition, we derive a new set of formulae to describe the consistency relations of low modes incorporating the effect of standing waves, which are evaluated and validated using model output. Finally, we conclude by providing a general discussion on the wave interference and its impact on the consistency relations.
\section{Regional Model}
\label{sec:modobs}
We continue to analyze the output of the regional MITgcm simulation considered in \cite{nelson2020improved} and subsequently in \cite{pan2020numerical,thakur2022impact,skitka2024internal,skitka2024probing,momeni2024breaking}. The simulation covers the Northeast Pacific Ocean between 24$^\circ$N and 32$^\circ$N and between 193$^\circ$E and 199$^\circ$E (Fig.\,\ref{fig:domain}), with a total time duration of 80 days. We focus on the higher-resolution simulation between the two discussed in~\cite{skitka2024internal} that employs a uniform horizontal grid spacing of $250~\text{m}$ and a vertical grid featuring 264 stretched vertical layers, with thicknesses ranging from $1~\text{m}$ near the surface to $25~\text{m}$ near and below the 300-m depth. The lateral boundary conditions are obtained from the hydrostatic global ocean model, LLC4320, which has a nominal horizontal grid spacing of $1/48^{\circ}$ and includes remotely-generated IGWs~\citep{nelson2020improved}. Simulated velocity and density fields are extracted for a 5-day period with a sampling interval of $500~\text{s}$.

Having produced Fig.\,\ref{fig:r1and2} from the model output, we now proceed with the reasoning in Section~\ref{sec:intro} that the established evaluation procedure of the consistency relations should be expected to only hold for high modes. Therefore, we propose to test Eqs.\,\eqref{eq:r11} and~\eqref{eq:r22} in a field where the low vertical modes are filtered out. To perform such an analysis, we first evaluate the 2D energy spectrum $E_*(k,\omega)$ on a fixed horizontal plane, where $k=\sqrt{k_x^2+k_y^2}$ with $k_x$ and $k_y$ being the zonal and meridional wave numbers, and $*$ represents $hk$, $vk$ or $p$. We then define a threshold vertical wavenumber $m_c$ and the energy spectrum with the low modes filtered out as

\begin{equation}
    E_*^{m_c}(k,\omega)= 
     \begin{cases}
     E_*(k,\omega)  &\text{if}~|m|~>~m_c \\
     0 &\text{if}~|m|~\le~m_c
  \end{cases},
\end{equation}
where the vertical wavenumber $m$ is evaluated based on the dispersion relationship 
\begin{eqnarray}
    m=\pm k\sqrt{\frac{N^2}{\omega^2-f^2}}.
\end{eqnarray}
Finally, the 1D frequency spectrum is evaluated as
\begin{equation}
    E^{m_c}_*(\omega)=\int E_*^{m_c}(k,\omega) dk.
\end{equation}

\begin{figure}
  \centerline{\includegraphics[trim=0cm 0cm 0cm 0.0cm, clip,scale=0.75]{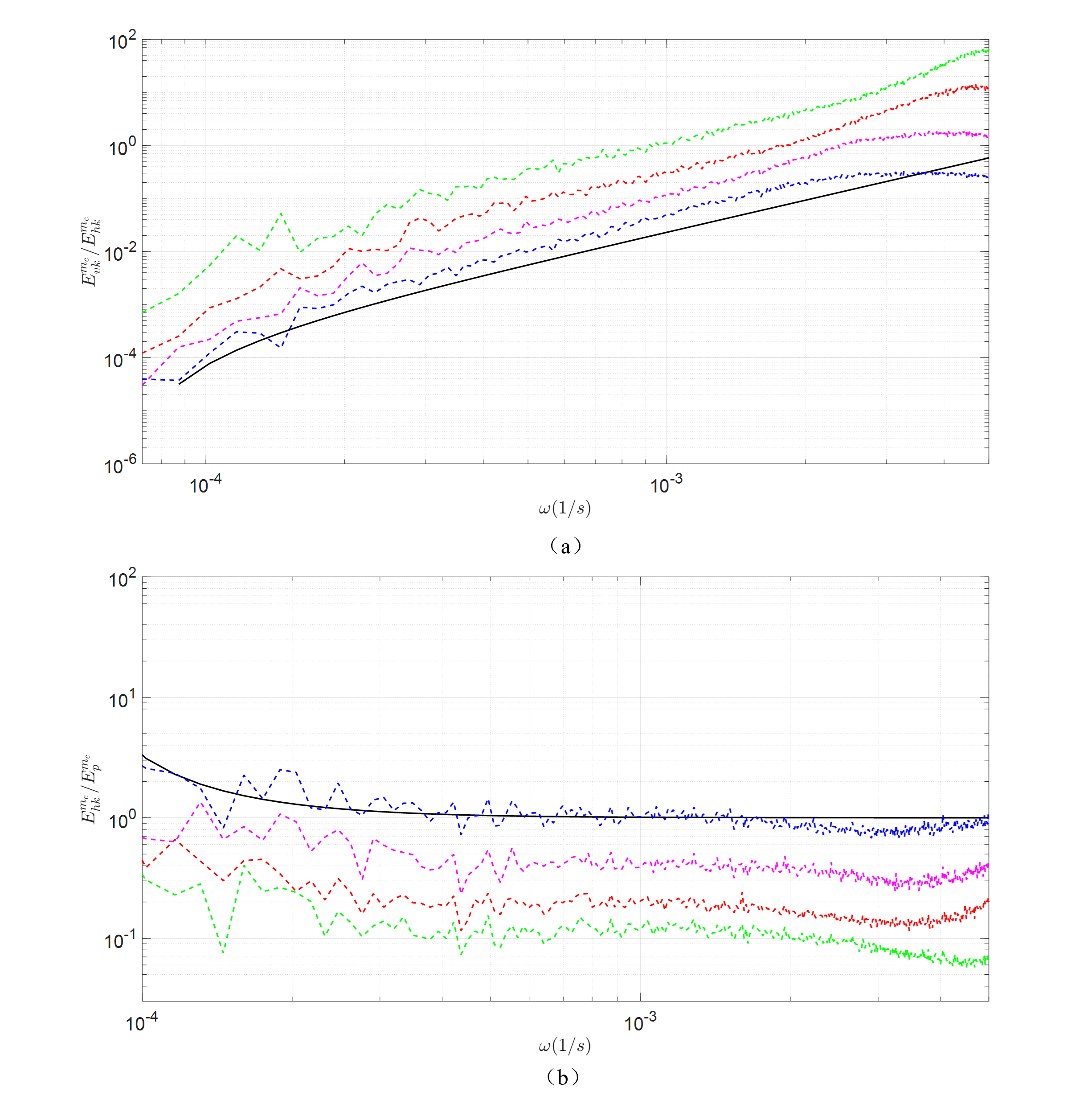}}
  \caption{Results of (a) $E^{m_c}_{vk}/E^{m_c}_{hk}$ and (b) $E^{m_c}_{hk}/E^{m_c}_{p}$ with $m_c=0$ (green dashed line), $m_c=0.002~\text{m}^{-1}$ (red dashed line), $m_c=0.004~\text{m}^{-1}$ (magenta dashed line), $m_c=0.008~\text{m}^{-1}$ (blue dashed line) for the depth of $4987~\text{m}$, in comparison with Eqs.\,\eqref{eq:r11} and \eqref{eq:r22} (black solid line), respectively.}
\label{fig:r1and2filtering}
\end{figure}

Fig.\,\ref{fig:r1and2filtering} presents the ratios, ${E^{m_c}_{vk}(\omega)}/{E^{m_c}_{hk}(\omega)}$ and ${E^{m_c}_{hk}(\omega)}/{E^{m_c}_{p}(\omega)}$, for $m_c=0, 0.002, 0.004$, and $0.008~\text{m}^{-1}$ at a depth of $4987~\text{m}$. It is shown that as $m_c$ increases (meaning more low modes are filtered out), the model results progressively align better with the classical consistency relations. Notably, when $m_c=0.008~\text{m}^{-1}$, the magnitudes from the model are very close to the theoretical values in both figures. This confirms our reasoning that the existence of low vertical modes is the culprit underlying deviation from Eqs.\,\eqref{eq:r11} and~\eqref{eq:r22} when the established classical procedure of evaluation is conducted. Because these low modes contain most energy of the IGW field, further investigation and a more robust theoretical framework are desired to fully understand this phenomenon.

\section{New Formula of Consistency Relations}
\label{sec:newcon}
\subsection{Derivation}
We begin by reviewing the derivation of classical consistency relations for IGWs, e.g., Eqs.\,\eqref{eq:r11} and~\eqref{eq:r22}. The linearized motion of disturbance in a rotational and stratified fluid with hydrostatic approximation can be described as
\begin{equation} 
\frac{\partial u}{\partial t}-fv=-\frac{\partial \pi}{\partial x},
\label{eq:g1}
\end{equation}

\begin{equation} 
\frac{\partial v}{\partial t}+fu=-\frac{\partial \pi}{\partial y},
\end{equation}

\begin{equation} 
\frac{\partial w}{\partial t}=-\frac{\partial \pi}{\partial z}-b,
\end{equation}

\begin{equation} 
\frac{\partial u}{\partial x}+\frac{\partial v}{\partial y}+\frac{\partial w}{\partial z}=0,
\label{eq:g5}
\end{equation}

\begin{equation} 
\frac{\partial \rho}{\partial t}+w\frac{\partial \bar{\rho}}{\partial z}=0,
\label{eq:g6}
\end{equation}
where $\pi\equiv {p}/{\rho_0}$ represents the kinematic pressure with $p$ the pressure. Assuming plane progressive waves, the solution to Eqs.\,\eqref{eq:g1}-\eqref{eq:g6} can be readily obtained as,
\begin{equation}
    u=\sqrt{\frac{k^2}{m^2|\boldsymbol{q}|^2}}\frac{m^2}{k^2} \sqrt{k_x^2+(fk_y/\omega)^2}\cos{(mz+k_xx+k_yy{+}\omega t+\theta_u)},
\label{eq:u1}
\end{equation}

\begin{equation}
    v=\sqrt{\frac{k^2}{m^2|\boldsymbol{q}|^2}} \frac{m^2}{k^2} \sqrt{k_y^2+(fk_x/\omega)^2}\cos{(mz+k_xx+k_yy{+}\omega t+\theta_v)},
\label{eq:v1}
\end{equation}

\begin{equation}
    w=-\sqrt{\frac{k^2}{m^2|\boldsymbol{q}|^2}}m\cos{(mz+k_xx+k_yy{+}\omega t)},
\label{eq:w1}
\end{equation}

\begin{equation}
    \pi=-\sqrt{\frac{k^2}{m^2|\boldsymbol{q}|^2}}\frac{N^2-\omega^2}{\omega} \cos{(mz+k_xx+k_yy{+}\omega t)},
\label{eq:pi1}
\end{equation}

\begin{equation}
    b=-\sqrt{\frac{k^2}{m^2|\boldsymbol{q}|^2}}\frac{m N^2}{\omega} \sin{(mz+k_xx+k_yy{+}\omega t)},
\label{eq:b1}
\end{equation}
with dispersion relation
\begin{equation}
    \omega^2=f^2+N^2\left(\frac{k}{m}\right)^2,
\label{eq:disp}
\end{equation}
where $\boldsymbol{q}=(k_x,k_y,m)$ is the three-dimensional wave vector, and $\theta_u=-\arctan\left({fk_y}/{\omega k_x}\right)$ and $\theta_v=\arctan\left({fk_x}/{\omega k_y}\right)$ are the phase angles.
The classical consistency relations can then be obtained by examining the amplitudes in Eqs.\,\eqref{eq:u1}-\eqref{eq:b1}, e.g., Eqs.\,\eqref{eq:r11} and~\eqref{eq:r22} obtained by substituting the amplitudes of Eqs.\,\eqref{eq:u1}-\eqref{eq:w1} and~\eqref{eq:b1} into Eqs.\,\eqref{eq:vkhk}-\eqref{eq:pe} for the respective energy ratios. 

\begin{figure}
  \centerline{\includegraphics[trim=0cm 0cm 0cm 0.5cm, clip,scale=0.5]{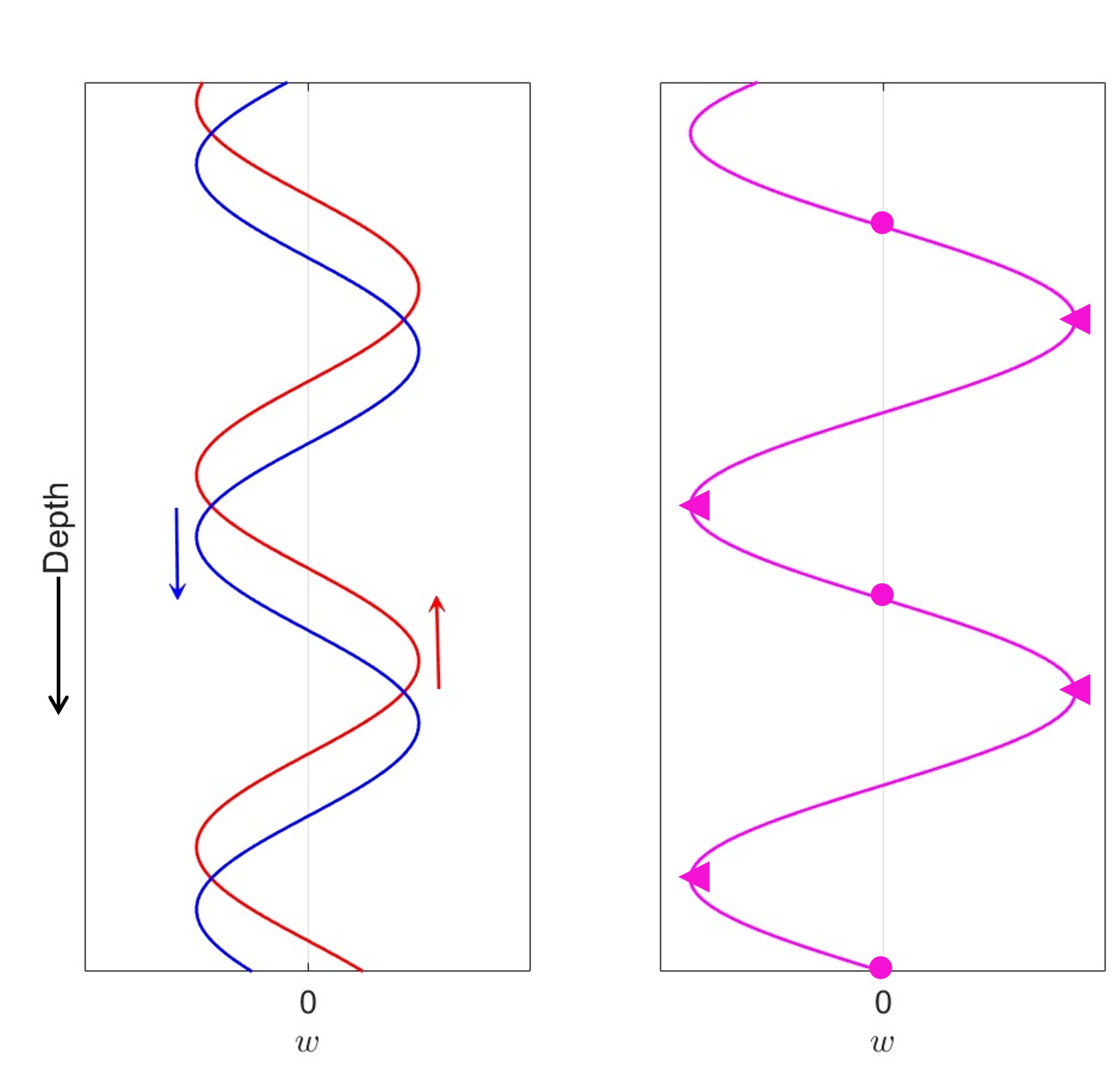}}
  \caption{A schematic drawing of the incident (blue solid line) and reflected (red solid line) waves (left), as well as the resulting standing wave (magenta solid line, right) with nodes (magenta dots) and antinodes (magenta triangles)}
\label{fig:standing_wave}
\end{figure}

We further note that any superposition of modes described by Eqs.\,\eqref{eq:u1}-\eqref{eq:b1} is also a solution to Eqs.\,\eqref{eq:g1}-\eqref{eq:g6}, due to the their linear nature. However, certain superpositions can break the spatial homogeneity of the wave field, and therefore affect the evaluation of the consistency relations. Let us now consider such a scenario in the presence of the ocean bottom, which reflects the incident wave and generates a superposed wave field.  Fig.\,\ref{fig:standing_wave} shows a schematic demonstration of the incident, reflected, and superposed total wave fields. Considering the vertical velocity field, if the solution for the incident wave is expressed in Eq.\,\eqref{eq:w1}, then the reflected wave needs to be formulated as 

\begin{equation}
    w^r=\sqrt{\frac{k^2}{m^2|\boldsymbol{q}|^2}}m \cos{\left[mz-(k_xx+k_yy{+}\omega t)\right]},
\label{eq:w2}
\end{equation}
so that the superposed wave field
\begin{equation}
    w^t=w+w^r=2\sqrt{\frac{k^2}{m^2|\boldsymbol{q}|^2}}m \sin(mz)\sin (k_xx+k_yy+\omega t)
\label{eq:wt}
\end{equation} 
satisfies the non-penetration boundary condition at the bottom $z=0$. From Eq.\,\eqref{eq:wt} we see that the total wave field is in the form of a standing wave in the vertical direction and progressive wave in the horizontal direction. Moreover, nodes and anti-nodes are formed at particular depths depending on the vertical wavenumber of the mode. It is clear now that such a standing wave pattern affects the evaluation of the consistency relation Eq.\,\eqref{eq:r11}, e.g., if a time series is sampled at a node where $w^t=0$, then $E_{vk}=0$, which clearly violates the relation. 

In accordance with \eqref{eq:w2} and \eqref{eq:wt}, other superposed fields can be formulated as
\begin{equation}
    u^t= 2 \sqrt{\frac{k^2}{m^2|\boldsymbol{q}|^2}} \frac{m^2}{k^2} \sqrt{k_x^2+(fk_y/\omega)^2}\cos(mz) \cos(k_xx+k_yy+\omega t+\theta_u),
\label{eq:u2}
\end{equation}
\begin{equation}
        v^t =2 \sqrt{\frac{k^2}{m^2|\boldsymbol{q}|^2}} \frac{m^2}{k^2} \sqrt{k_y^2+(fk_x/\omega)^2} \cos(mz) \cos (k_xx+k_yy+\omega t+\theta_v),
\label{eq:v2}
\end{equation}

\begin{equation}
    b^t=-2 \sqrt{\frac{k^2}{m^2|\boldsymbol{q}|^2}}\frac{m N^2}{\omega}\sin(mz)\cos (k_xx+k_yy+\omega t),
\label{eq:b2}
\end{equation}

\begin{equation}
    \pi^t=-2 \sqrt{\frac{k^2}{m^2|\boldsymbol{q}|^2}}\frac{N^2-\omega^2}{\omega}\cos(mz)\cos(k_xx+k_yy+\omega t).
\label{eq:p2}
\end{equation}

A new set of consistency relations considering the ocean bottom can then be constructed by examining the amplitudes in \eqref{eq:wt}-\eqref{eq:p2}. The counterparts of Eqs.\,\eqref{eq:r11} and~\eqref{eq:r22} can now be expressed as
\begin{eqnarray}
    \frac{E_{vk}}{E_{hk}}=\frac{(\omega^2-f^2)\omega^2}{(\omega^2+f^2)N^2}{\tan^2(mz)},
\label{eq:r1new}
\end{eqnarray}

\begin{equation}
    \frac{E_{hk}}{E_p}=\frac{\omega^2+f^2}{\omega^2-f^2}{\tan^{-2}(mz)}.
\label{eq:r2new}
\end{equation}
Before proceeding to the next section for validation, a few comments regarding Eqs.~\eqref{eq:r1new} and~\eqref{eq:r2new} are warranted. First, we note that the new formulae are established for certain vertical modes with wavenumber $m$, i.e., they describe the energy ratios for such modes as frequency (or horizontal wavenumber) varies. Second, while the free surface serves as another vertical boundary, its mechanism is different from that of the bottom. Instead of imposing a no-penetration boundary condition, the surface deforms in a way that the surface vertical velocity is equal to $w$. Such a mechanism turns out to have no impact on the consistency relations as illustrated in Section 1.  

\subsection{Results and Validation}
In this section, we validate the formulae ~\eqref{eq:r1new}-\eqref{eq:r2new} using output of the regional model described in Section~\ref{sec:modobs}. We focus on those low vertical modes for which deviations from Eqs.\,\eqref{eq:r11}-\eqref{eq:r22} have been presented. In order to validate Eqs.\,\eqref{eq:r1new}-\eqref{eq:r2new}, it is necessary to compute the spectrum $E_*(m,\omega)$. For this purpose we choose to first compute the spectrum $E_*(k,\omega)$ which can be easily obtained from data on a horizontal plane within a specified time interval. Then $E_*(m,\omega)$ is obtained by
\begin{equation}
    E_*(m,\omega)=E_*(k,\omega)\frac{dk}{dm},
\end{equation}
with
\begin{equation}
    \frac{dk}{dm}=\sqrt{\frac{\omega^2-f^2}{N^2}}.
\end{equation}
We note that in the conversion from the space of $(k,\omega)$ to $(m,\omega)$, the numerical grids for $k$ and $m$ cannot be both uniformly spaced, so interpolation is inevitably needed here. In particular, we first pick up the values of interest for $m$, then determine the values of $k$ corresponding to $(m,\omega)$, and finally the interpolation is performed in the space of $k$.

\begin{figure}
  \centerline{\includegraphics[trim=0cm 0cm 0cm 0.0cm, clip,scale=0.75]{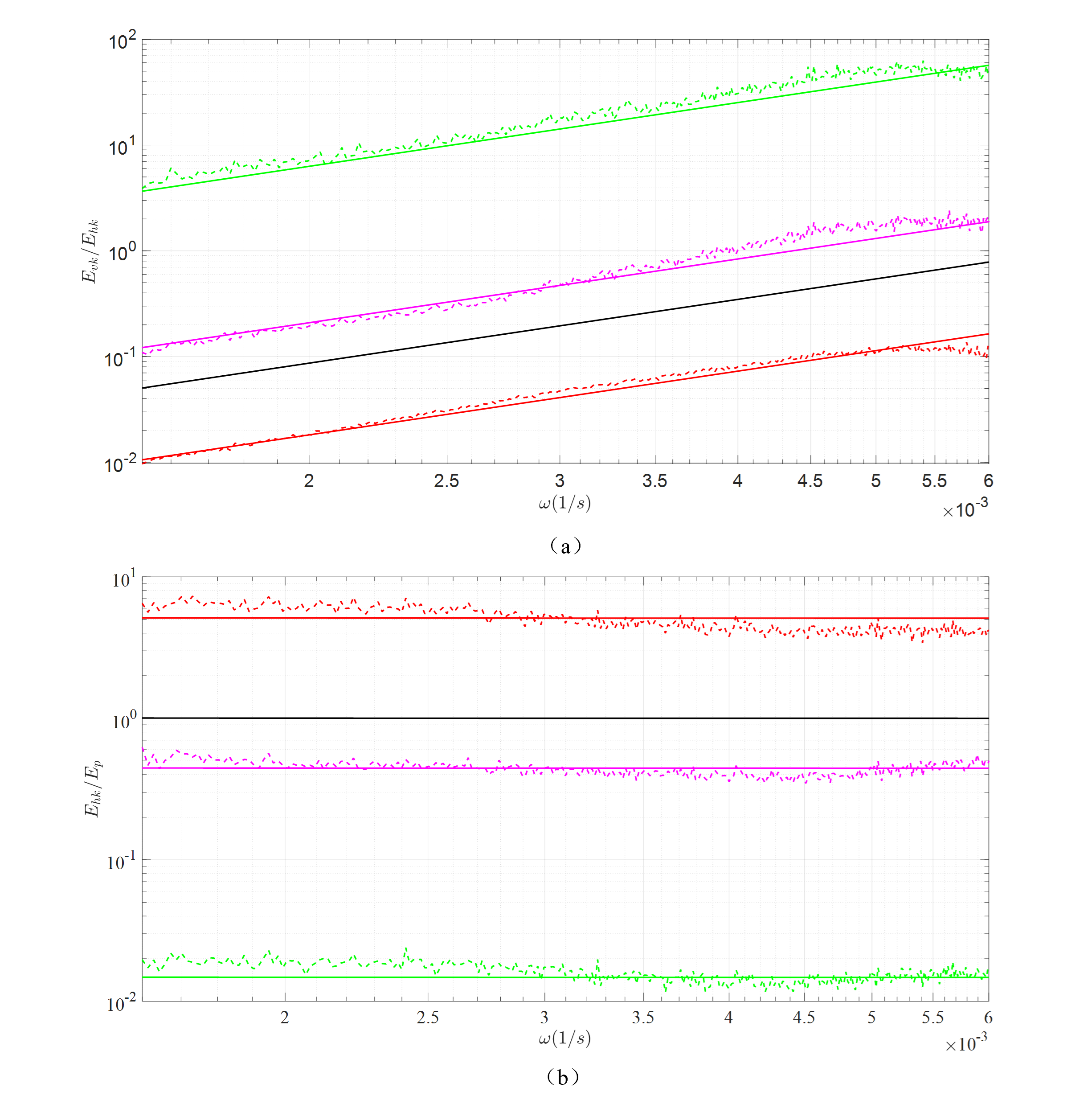}}
  \caption{Results of (a) $E_{vk}/E_{hk}$ (b) $E_{hk}/E_{p}$ obtained from the model and Eqs.\,\eqref{eq:r1new}-\eqref{eq:r2new} at the heights of $3350~\text{m}$ (red dashed and solid lines), $2650~\text{m}$ (magenta dashed and solid lines) and $725~\text{m}$ (green dashed and solid lines) with $m=0.002~\text{m}^{-1}$, in comparison with Eqs.\,\eqref{eq:r11} and \eqref{eq:r22} (black solid line), respectively}
\label{fig:r1and2_002}
\end{figure}

\begin{figure}
  \centerline{\includegraphics[trim=0cm 0cm 0cm 0.0cm, clip,scale=0.75]{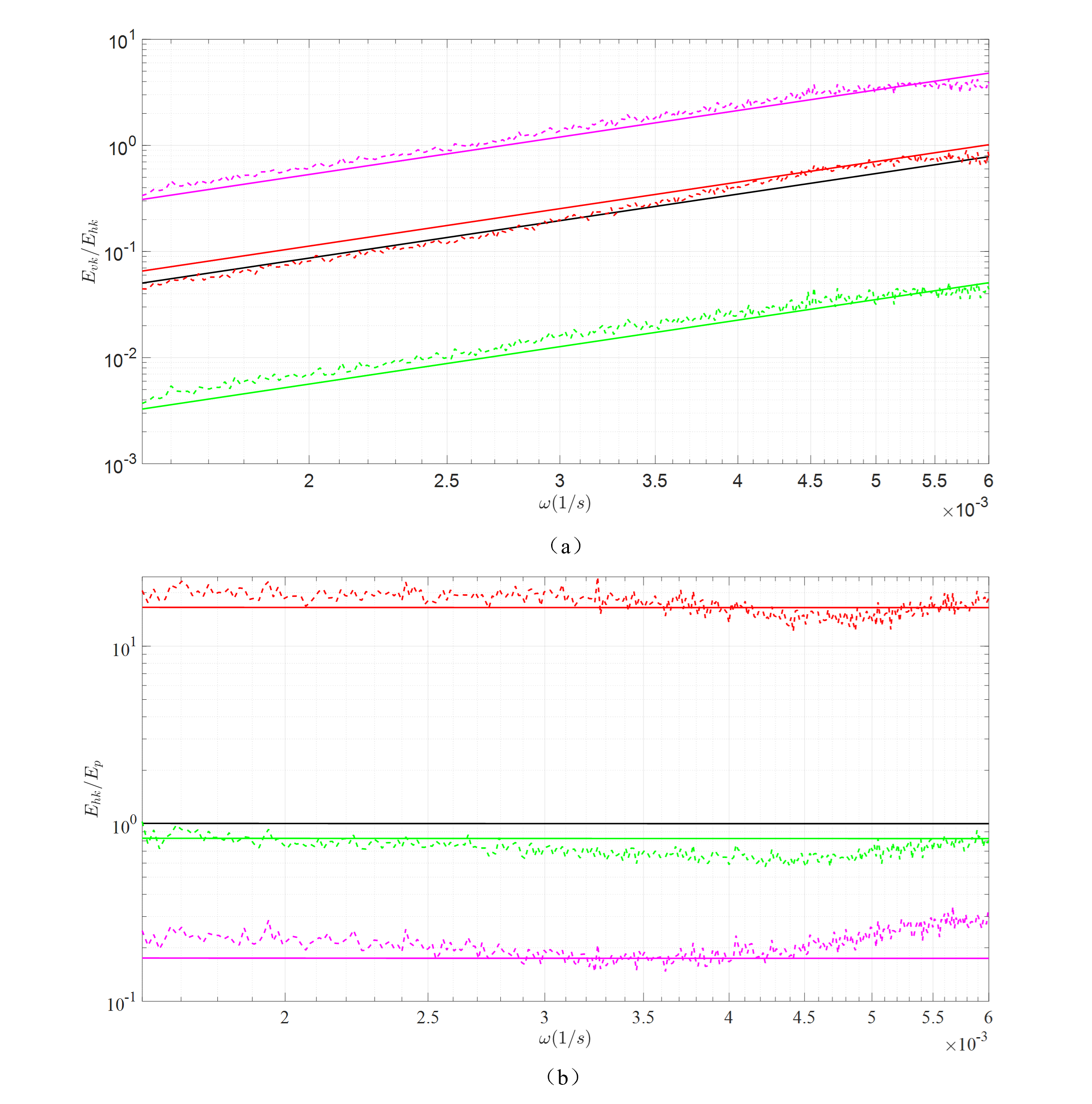}}
  \caption{Results of (a) $E_{vk}/E_{hk}$ (b) $E_{hk}/E_{p}$ obtained from the model and Eqs.\,\eqref{eq:r1new}-\eqref{eq:r2new} at the heights of $3350~\text{m}$ (red dashed and solid lines), $2650~\text{m}$ (magenta dashed and solid lines) and $725~\text{m}$ (green dashed and solid lines) with $m=0.004~\text{m}^{-1}$, in comparison with Eqs.\,\eqref{eq:r11} and \eqref{eq:r22} (black solid line), respectively}
\label{fig:r1and2_004}
\end{figure}

We now test the formulae~\eqref{eq:r1new}-\eqref{eq:r2new} at three different depths of $2362~\text{m}$, $3062~\text{m}$ and $4987~\text{m}$, corresponding to heights from the sea bottom of $3350~\text{m}$, $2650~\text{m}$, and $725~\text{m}$. Fig.\,\ref{fig:r1and2_002} presents the results of ${E_{vk}}/{E_{hk}}$ and ${E_{hk}}/{E_{p}}$ as functions of $\omega$ with $m=0.002~\text{m}^{-1}$, in comparison with the new formulae Eqs.\,\eqref{eq:r1new}-\eqref{eq:r2new} and the classical consistency relations Eqs.\,\eqref{eq:r11} and~\eqref{eq:r22}. It can be found that the model results agree well with Eqs.\,\eqref{eq:r1new}-\eqref{eq:r2new}, but not Eqs.\,\eqref{eq:r11}-\eqref{eq:r22}, at all three depths. A similar analysis has also been performed for $m=0.004~\text{m}^{-1}$, with the results shown in Fig.\,\ref{fig:r1and2_004} further supporting the validity of Eqs.\,\eqref{eq:r1new}-\eqref{eq:r2new}. We note that for values of $m$ above the two choices here, the dynamics is described well by the classical consistency relation as illustrated in Fig.\,\ref{fig:r1and2filtering}

\begin{figure}
  \centerline{\includegraphics[trim=0cm 0cm 0cm 0.0cm, clip,scale=0.75]{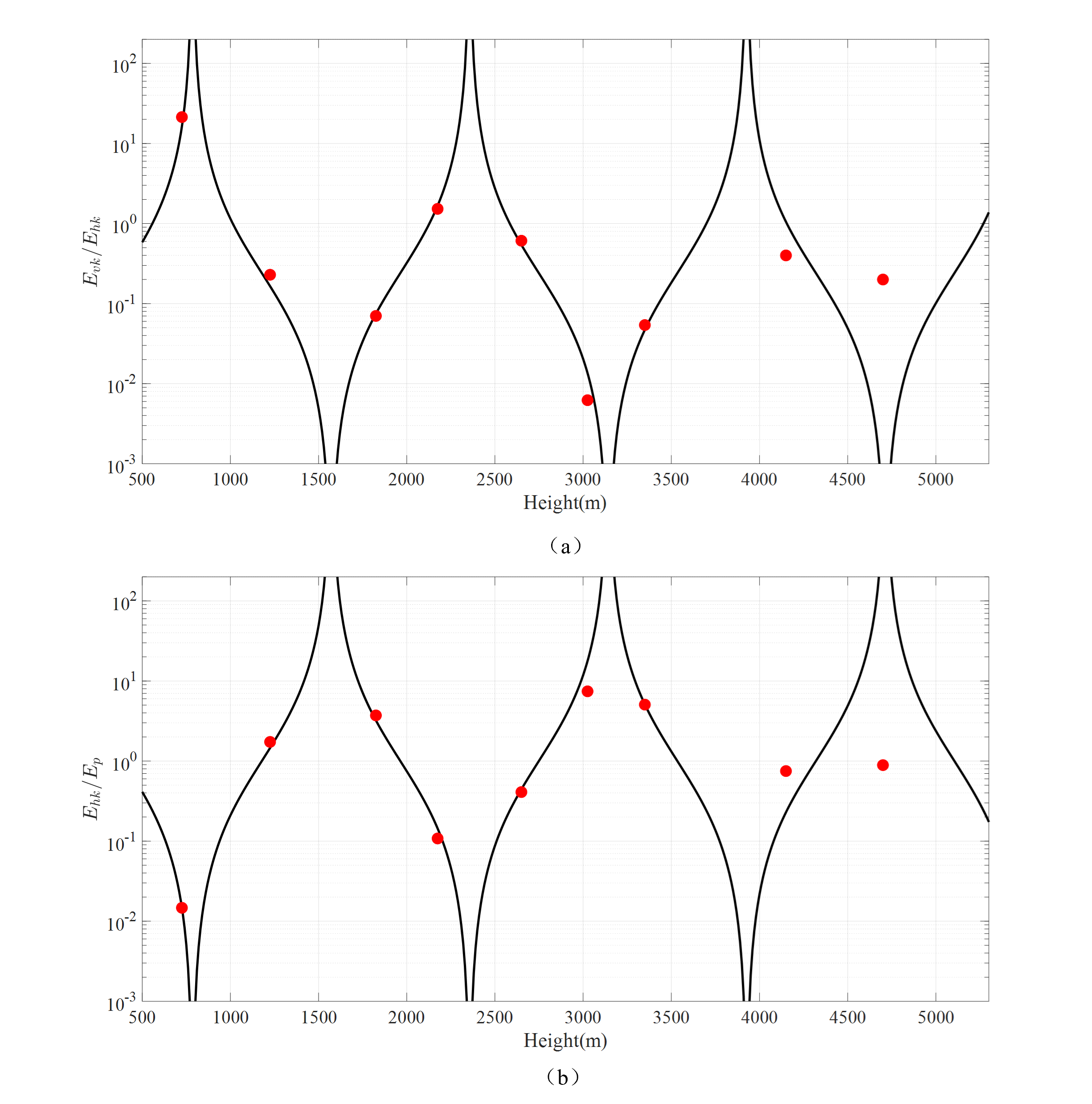}}
  \caption{Results of (a) $E_{vk}/E_{hk}$ and  (b) $E_{hk}/E_{p}$ at different heights obtained from Eqs.\,\eqref{eq:r1new} and \eqref{eq:r2new} (black solid line), respectively, and the model (red dots), with $m=0.002~\text{m}^{-1}$ and $\omega=3.18\times 10^{-3}~\text{s}^{-1}$}
\label{fig:r_height_002}
\end{figure}

\begin{figure}
  \centerline{\includegraphics[trim=0cm 0cm 0cm 0.0cm, clip,scale=0.75]{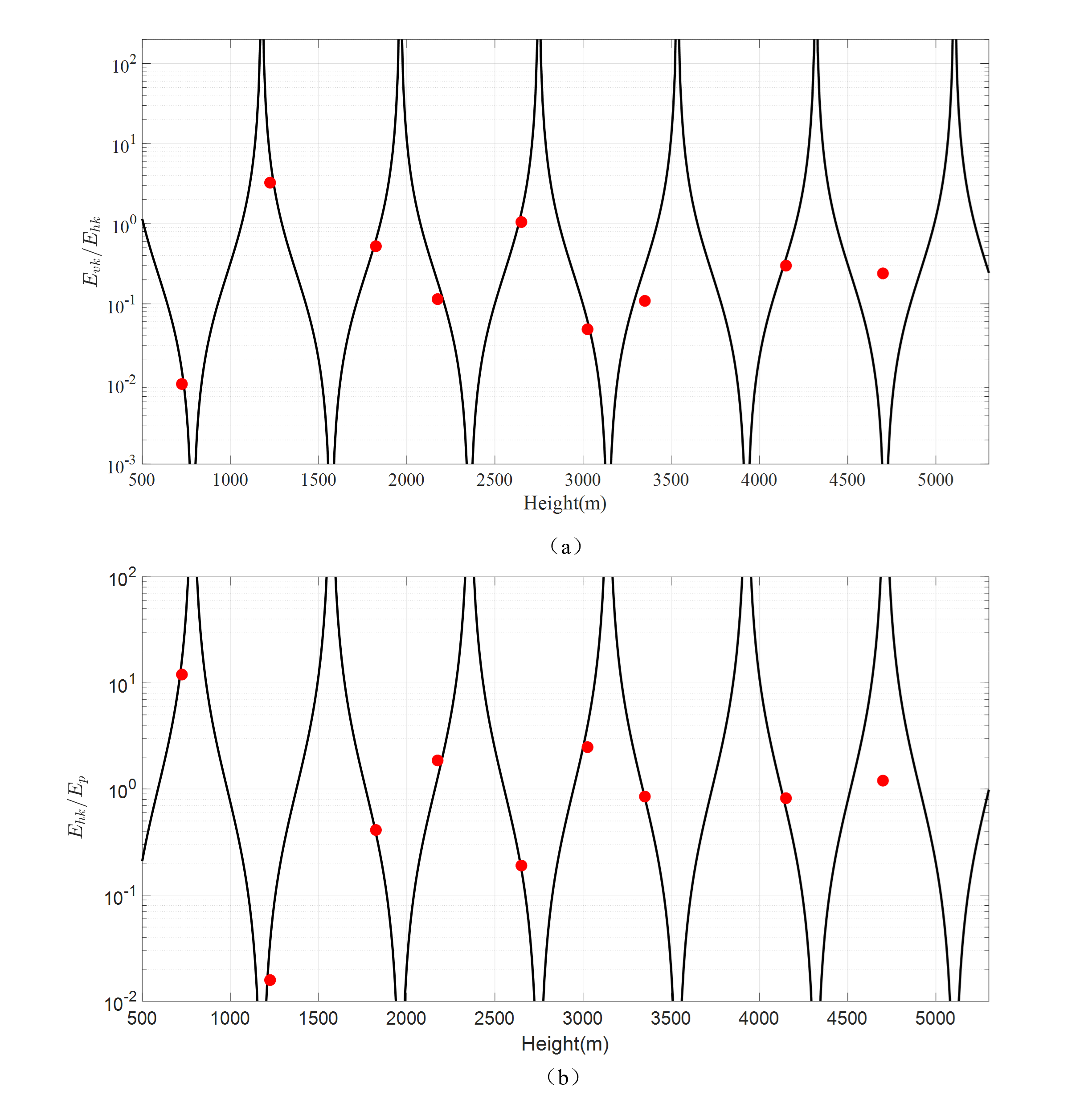}}
  \caption{Results of (a) $E_{vk}/E_{hk}$ and  (b) $E_{hk}/E_{p}$ at different heights obtained from Eqs.\,\eqref{eq:r1new} and \eqref{eq:r2new} (black solid line), respectively, and the model (red dots), with $m=0.004~\text{m}^{-1}$ and $\omega=3.18\times 10^{-3}~\text{s}^{-1}$}
\label{fig:r_height_004}
\end{figure}



We next test more explicitly the dependence on $z$ of the new formulae~\eqref{eq:r1new}-\eqref{eq:r2new}. For this purpose, we fix the value of $m$ (to be either $m=0.002~\text{m}^{-1}$ or $m=0.004~\text{m}^{-1}$) and frequency $\omega=3.18\times 10^{-3}~\text{s}^{-1}$. The results of ${E_{vk}}/{E_{hk}}$ and ${E_{hk}}/{E_{p}}$ at $9$ different heights (including $725~\text{m}$, $1225~\text{m}$, $1825~\text{m}$, $2175~\text{m}$, $2575~\text{m}$, $3025~\text{m}$, $3350~\text{m}$, $4150~\text{m}$, and $4700~\text{m}$) from the model are plotted in Figs.~\ref{fig:r_height_002}-\ref{fig:r_height_004} for the two choices of $m$. We see that for depth levels closer to the bottom (i.e., heights less than 3350\text{m}), the model results agree with Eqs.\,\eqref{eq:r1new}-\eqref{eq:r2new} in terms of the tangent-style periodic pattern. For higher levels, the model data lose the tangent-style pattern as expected, reflecting the physics of a diminished effect of the ocean bottom.

\section{Discussion}

In Sections~\ref{sec:modobs} and~\ref{sec:newcon}, we have demonstrated that the wave interference of low vertical modes near the bottom can cause deviation from the consistency relations, given that $E_*(\omega)$ is evaluated from the time series taken from a single level close to the bottom. However, there are still a few more subtle issues we need to address before concluding. In particular, wave interference exists everywhere in the ocean interior. It is foreseeably true that the interference of low vertical modes near the bottom is sufficiently strong due to the standing wave pattern as discussed in Section~\ref{sec:newcon}. However, further investigation is needed to understand the potential impact of interference at other locations within the ocean interior and from a large number of wave modes propagating in different directions (e.g., on a horizontal plane). Specifically, it remains to be investigated whether such interference can induce appreciable deviations from the classical consistency relations, e.g., Eqs.~\eqref{eq:r11} and~\eqref{eq:r22}.

\begin{figure}
  \centerline{\includegraphics[trim=0cm 0cm 0cm 0.0cm, clip,scale=0.4]{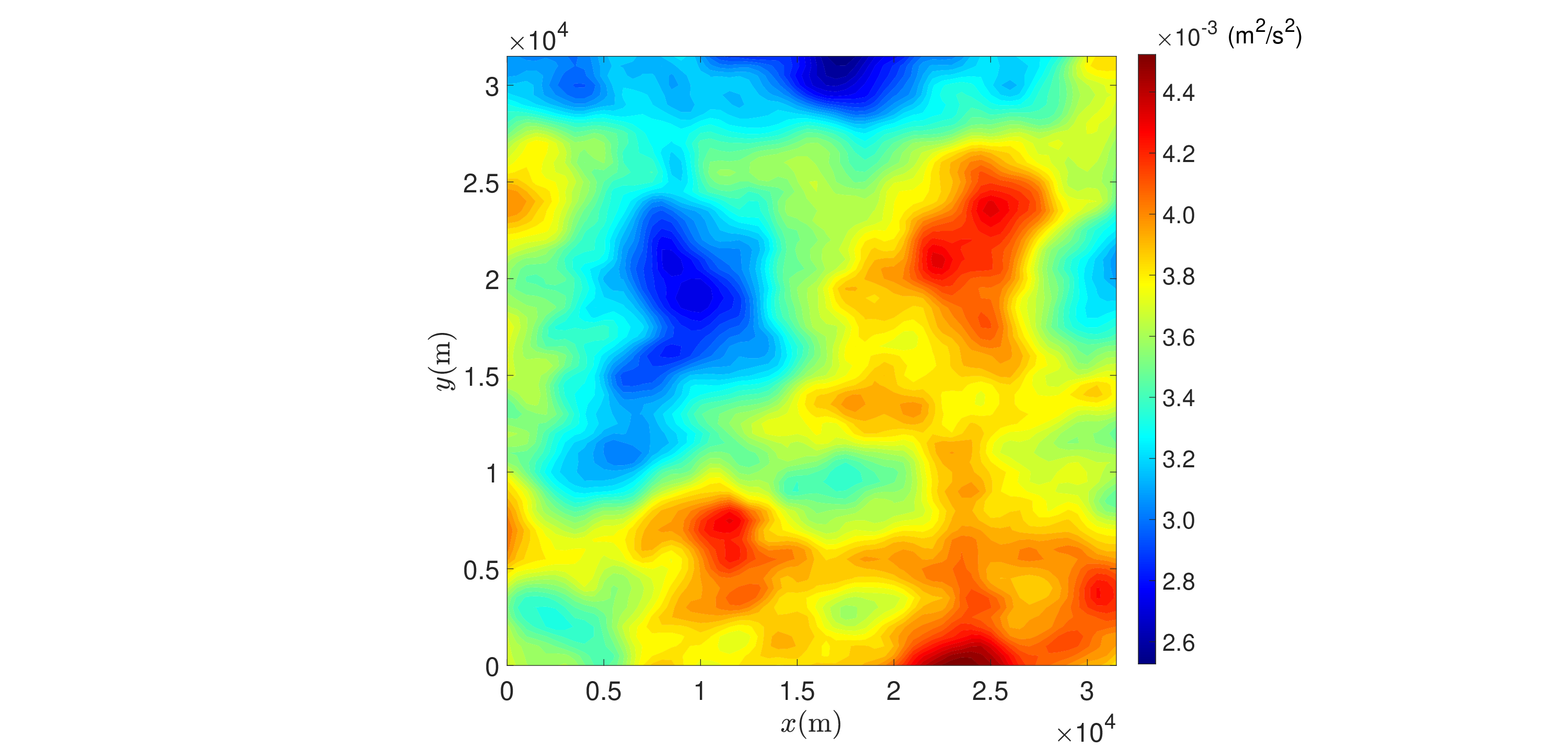}}
  \caption{$\mathcal{E}_{hk}(x,y)$ at a fixed depth of $3062~\text{m}$.}
\label{fig:HKEXY}
\end{figure}

We begin by discussing the interference of waves on a horizontal plane. As shown in \cite{pan2020numerical} (see Fig.\,7 there), these waves are mostly isotropic, i.e., for a given frequency, waves are present in all horizontal directions with comparable magnitudes. Intuitively speaking, when considering the sum of many such modes, all with random phases, the impact of interference may tend to average out or at least not be sufficiently strong. We can confirm this point directly based on the model output. Fig.\,\ref{fig:HKEXY} plots $\mathcal{E}_{hk}(x,y)=\int_f^{\infty}E_{hk}(\omega,x,y)d\omega$, where $E_{hk}(\omega,x,y)$ is the horizontal kinetic energy frequency spectrum evaluated based on the time series taken at a spatial point $(x,y)$ with a fixed depth of $3062~\text{m}$. Within the examined area of $32~\text{km}\times32~\text{km}$, the variation of $\mathcal{E}_{hk}(x,y)$ due to the horizontal interference is below $50\%$ as seen in Fig.\,\ref{fig:HKEXY}. Such a level of variation is therefore not enough to induce remarkable deviations from the consistency relations when plotted in logarithmic scale as in Fig.\,\ref{fig:r1and2}.

\begin{figure}
  \centerline{\includegraphics[trim=0cm 0cm 0cm 0.0cm, clip,scale=0.3]{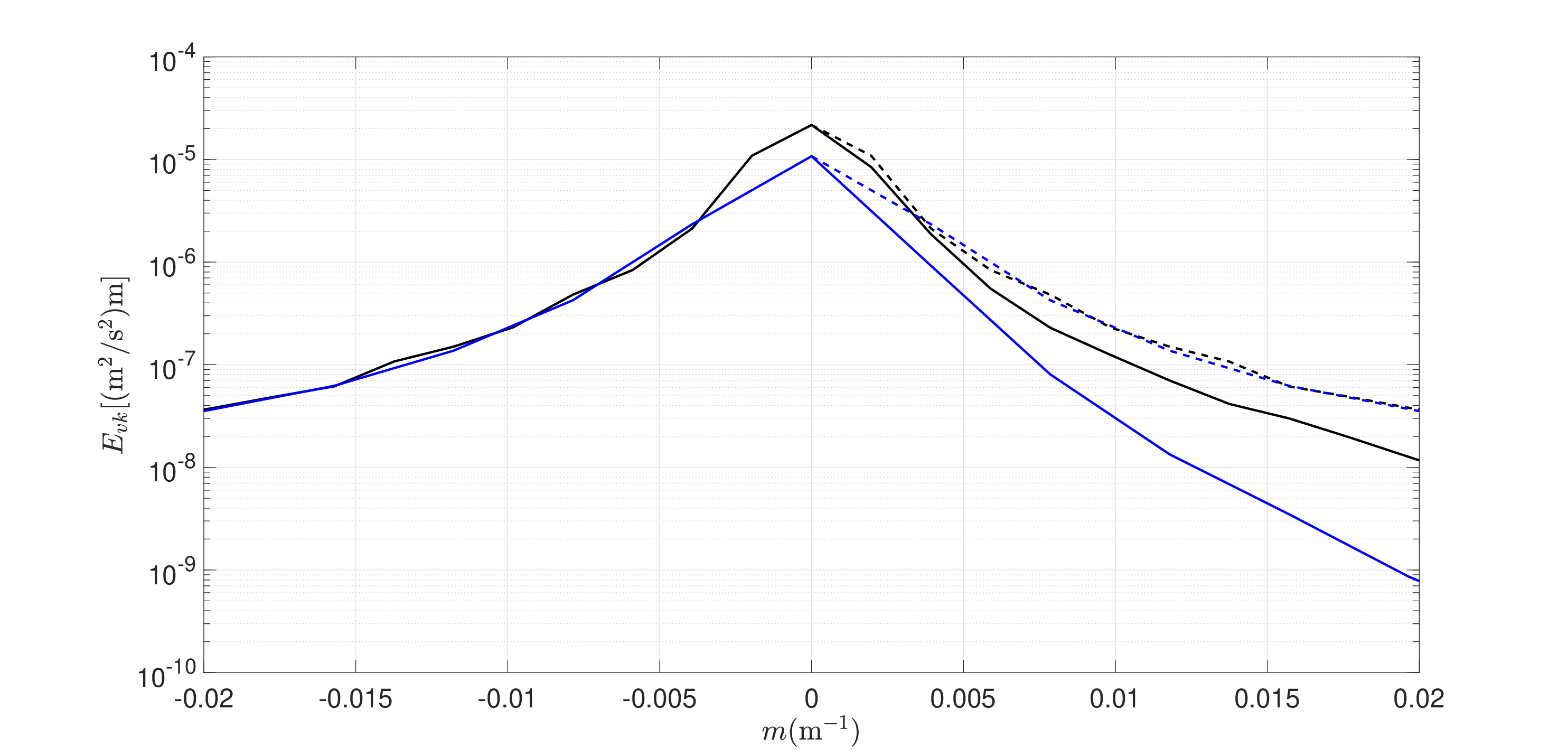}}
  \caption{$E_{vk}(m)$ obtained in $\mathcal{Z}_1=[500~\text{m}, 3750~\text{m}]$ (black solid line) and $\mathcal{Z}_2=[4050~\text{m}, 5627~\text{m}]$ ({blue solid line}). The black and blue dashed lines are produced by mirroring $E_{vk}(m)$ of $m<0$ with respect to $m=0$ to illustrate the level of symmetry of the spectrum.}
\label{fig:m_spectrum_two_ranges}
\end{figure}

We next discuss the interference of vertical modes at different depths. The results we show in Fig.\,\ref{fig:r1and2} imply that such interference of low vertical modes is strong only near the bottom. However, at any ocean depth, there are both upward and downward propagating waves. In order for the interference to be strong only near the bottom, it has to be the case that the magnitudes of upward and downward propagating waves are comparable only near the bottom due to reflection, but differ more significantly at other depth levels. We confirm this point again using the model output. Specifically, we consider two height ranges, $\mathcal{Z}_1=[500~\text{m}, 3750~\text{m}]$ and $\mathcal{Z}_2=[4050~\text{m}, 5627~\text{m}]$, and compute $E_{vk}(m)=\int_f^{\infty}E_{vk}(m,\omega)d\omega$ (with $\omega\in \mathbb{R}^+$ and $m\in \mathbb{R}$) based on the model output of a specified vertical profile (i.e., the horizontal location is fixed). Under this setup, $E_{vk}(m)$ with $m>0$ and $m<0$ represents respectively waves traveling downward and upward. We present the results of $E_{vk}(m)$ for $-0.02~\text{m}^{-1}\leq m \leq 0.02~\text{m}^{-1}$ from both regions $\mathcal{Z}_1$ and $\mathcal{Z}_2$ in Fig.\,\ref{fig:m_spectrum_two_ranges}. It can be seen that the spectrum $E_{vk}(m)$ is much more symmetric with respect to $m=0$ when evaluated in $\mathcal{Z}_1$ than in $\mathcal{Z}_2$, supporting our earlier argument. In addition, we see that the symmetry tends to diminish with the increase of $m$ for both cases. This intuitively makes sense because with the decrease of wavelength, the effective distance from a vertical layer to the ocean bottom (defined as physical distance over the wavelength) increases so that the effect of ocean bottom on the dynamics is reduced.


\begin{figure}
  \centerline{\includegraphics[trim=0cm 0cm 0cm 0.0cm, clip,scale=0.75]{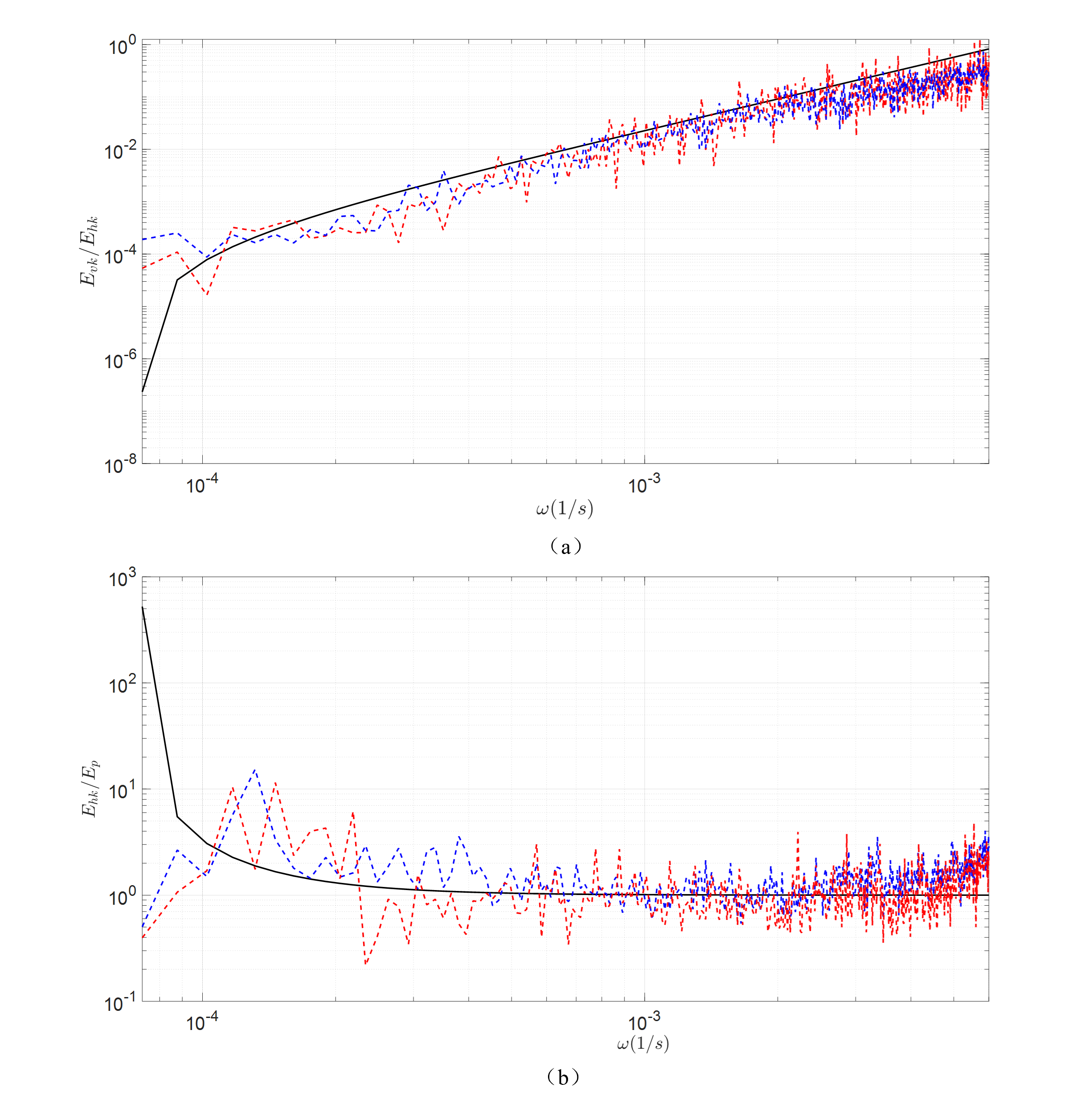}}
  \caption{(a) $E_{vk}/E_{hk}$ and (b) $E_{hk}/E_{p}$ for $m=-0.004~\text{m}^{-1}$ (red dashed line) and $m=-0.008~\text{m}^{-1}$ (blue dashed line), in comparison with Eqs.\,\eqref{eq:r11} and \eqref{eq:r22} (black solid line), respectively.}
\label{fig:single_mode}
\end{figure}

Finally, we remark that the deviation from the classical consistency relations discussed in this paper is caused by the way that $E_*$ is evaluated. The classical consistency relations are still well satisfied for each mode in the ocean. To confirm this, we calculate $E_*$ for only the upward propagating modes (respectively $m=-0.004~\text{m}^{-1}$ and $m=-0.008~\text{m}^{-1}$) from the region $\mathcal{Z}_1$ near the bottom, and plot $E_{vk}/E_{hk}$ and $E_{hk}/E_{p}$ with comparison to Eqs.\,\eqref{eq:r11} and~\eqref{eq:r22} in Fig.\,\ref{fig:single_mode}. It can be seen that for each single mode, the model results agree closely with Eqs.\,\eqref{eq:r11} and~\eqref{eq:r22} especially in the high-frequency regime. 

\section{Conclusions}
This paper begins with a review of the established methods to evaluate IGW activities through consistency relations. We point out that the results derived from time series at a fixed depth level or spatial point are fundamentally problematic when the method is applied to regions close to the ocean bottom. The problem lies in the existence of interference of low modes incident to and reflected from the ocean bottom. We accordingly derive a new set of formulae to describe the dynamics of the superposed low modes that are verified by the model output. Finally, we provide a general discussion on the interference of IGWs in the ocean and its potential impacts on consistency relations.


\section*{Acknowledgments}
All the simulations in this study were performed on the Niagara supercomputer, maintained by the SciNet facility, a component of the Digital Research Alliance of Canada located at the University of Toronto. G. W., Y. W., and Y. P. acknowledge funding from the National Science Foundation (AWD023422). B. K. A. acknowledges funding from NASA (80NSSC24K1649). The authors appreciate the insightful discussions with Dr. Kurt Polzin in the Woods Hole Oceanographic Institution.

    
    
    
    
    
    
    
    
    
    
    \bibliography{cas-refs}
    
    \end{document}